\documentclass{article}
\usepackage{ijcai25}

\usepackage{times}
\usepackage{soul}
\usepackage{url}
\usepackage[hidelinks]{hyperref}
\usepackage[utf8]{inputenc}
\usepackage[small]{caption}
\usepackage{graphicx}
\usepackage{amsmath}
\usepackage{amsthm}
\usepackage{booktabs}
\usepackage{algorithm}
\usepackage{algorithmic}
\usepackage[switch]{lineno}
\usepackage{amsmath,amssymb,amsfonts}  
\usepackage{enumitem}
\usepackage{booktabs, multirow}
\usepackage{mathtools}
\usepackage{changepage} 
\usepackage[margin=1in]{geometry} 
\newcommand{\emphasize}[1]{``#1''}
\usepackage{xcolor}

\newcommand{\hashimoto}[1]{\textcolor{black}{#1}}
\newcommand{\takehiro}[1]{\textcolor{black}{#1}}

\title{The Impact and Feasibility of Self-Confidence Shaping \\
for AI-Assisted Decision-Making}

\author{
Takehiro Takayanagi$^1$
\and
Ryuji Hashimoto$^1$\and
Chung-Chi Chen$^2$\And
Kiyoshi Izumi$^1$\\
\affiliations
$^1$The University of Tokyo\\
$^2$National Institute of Advanced Industrial Science and Technology\\
\emails
\{takayanagi-takehiro590, hashimoto-ryuji419\}@g.ecc.u-tokyo.ac.jp,
c.c.chen@acm.org,
izumi@sys.t.u-tokyo.ac.jp
}

\begin{document}

\maketitle

\begin{abstract}
In AI-assisted decision-making, it is crucial but challenging for humans to appropriately rely on AI, especially in high-stakes domains such as finance and healthcare.
This paper addresses this problem from a human-centered perspective by presenting an intervention for \textit{self-confidence shaping}, designed to calibrate self-confidence at a targeted level.
We first demonstrate the impact of self-confidence shaping by quantifying the upper-bound improvement in human-AI team performance. Our behavioral experiments with 121 participants show that self-confidence shaping can improve human-AI team performance by nearly 50\% by mitigating both over- and under-reliance on AI.\footnote{The experimental protocol was approved by our organization’s ethical board}
We then introduce a self-confidence prediction task to identify when our intervention is needed. Our results show that simple machine-learning models achieve 67\% accuracy in predicting self-confidence.
We further illustrate the feasibility of such interventions.
The observed relationship between sentiment and self-confidence suggests that modifying sentiment could be a viable strategy for shaping self-confidence.
Finally, we outline future research directions to support the deployment of self-confidence shaping in a real-world scenario for effective human-AI collaboration.
\end{abstract}

\section{Introduction}

Artificial Intelligence (AI) technologies have become ubiquitous and are adopted across diverse areas of society, leading to an increasing reliance on AI recommendations in decision-making. Yet, in high-stakes domains such as finance and healthcare, where safety, ethical, and legal obligations remain paramount, full automation can be problematic, while purely manual approaches risk inefficiency and inaccuracy. Consequently, AI-assisted decision-making has emerged as a promising paradigm, which augments human judgment with AI recommendations~\cite{lai2019human,towards_science,ijcai2024p868}.

A central challenge in this paradigm is ensuring appropriate reliance on AI~\cite{schemmer2023appropriate}. Over-reliance can degrade outcomes if the model or data are flawed, while under-reliance means ignoring beneficial advice. To address this, prior research has primarily focused on \textit{AI-centric interventions}, such as providing confidence scores, or explanations for AI predictions~\cite{yin2019understanding,effect_of_confidence,bansal2019beyond}.

\takehiro{Beyond AI-centric interventions, prior studies emphasize that human self-confidence is key to proper reliance.
However, human self-confidence is often treated as a fixed factor~\cite{mahmood2024designing,chong2022human}. This perspective overlooks the potential for a \textit{human-centric intervention} to shape human self-confidence and mitigate both over- and under-reliance. Consequently, this represents a largely underexplored avenue for improving AI-assisted decision-making.}
\begin{figure}[t]
    \centering
    \includegraphics[width=\linewidth]{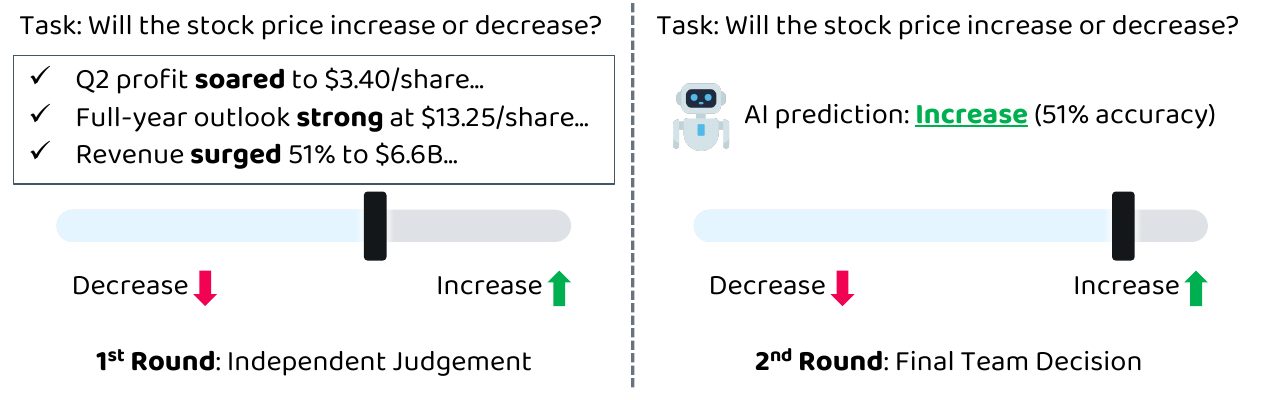}
\caption{%
    Conceptual illustration of the two-round procedure:
    \textbf{1\textsuperscript{st} round}, independent judgment; 
    \textbf{2\textsuperscript{nd} round}, final team decision with AI recommendation.
}
    \label{fig:task_illustration}
    \vspace{-5mm}
\end{figure}

In this paper, we take an initial step toward enhancing AI-assisted decision-making through \textit{self-confidence shaping}.\footnote{We define self-confidence shaping as the deliberate intervention to shape a decision-maker’s confidence at a targeted level. This differs from confidence calibration, which aligns a user’s confidence with their actual performance~\cite{moore2020perfectly}.} First, we quantify the potential gains in human-AI team performance under ideal conditions where self-confidence is perfectly controlled. Next, we introduce a self-confidence prediction task to determine when our intervention is needed, assessing how accurately a decision-maker's (DM) self-confidence can be predicted based on user traits and task characteristics. Finally, we explore the feasibility of this intervention by investigating the relationship between task features and self-confidence. In particular, drawing on decision science research that demonstrates how subtle shifts in sentiment in a text can significantly influence self-confidence~\cite{griffin1992weighing}, we examine how modifying sentiment may serve as an initial step toward effective self-confidence shaping.

To summarize, we explore the following questions.
\begin{itemize}
    \item \textbf{RQ1:} Under perfect control over self-confidence, how much can human-AI team performance improve through self-confidence shaping?
    \item \textbf{RQ2:} To what extent can we predict self-confidence in decision-making?
    \item \textbf{RQ3:} Is the sentiment significantly associated with self-confidence?
\end{itemize}

\takehiro{To address these questions, we developed an AI-assisted investment decision-making task in a financial context, recruited 121 crowd workers, and collected 2,420 decision instances. Adopting the Judge-Advisor System (JAS) for decision-making~\cite{sniezek1995cueing}, our experiments consist of two rounds: (1) participants first form an independent prediction, and (2) they make the final decision after considering the AI recommendation, as illustrated in Figure~\ref{fig:task_illustration}.
}

Our results show that successful self-confidence shaping can significantly improve human-AI team performance. In particular, our scenario analyses reveal that shaping self-confidence can reduce both over- and under-reliance on AI recommendations, improving team performance by up to nearly 50\%. Moreover, our findings suggest that such intervention is indeed feasible: simple machine learning models that incorporate user traits and task characteristics can predict whether self-confidence is high or low with about 67\% accuracy. We also observe that the sentiment in a text is significantly associated with self-confidence.

\section{Related Work}
\subsection{Appropriate Reliance in AI-assisted Decision-Making}
A challenge for improving human-AI team performance in AI-assisted decision-making is ensuring humans’ appropriate reliance on AI, given the uncertainties of both humans and AI~\cite{schemmer2023appropriate,does_the,buccinca2021trust,towards_science}.
Over-reliance on AI can degrade human-AI decision-making if the model is flawed or data is unreliable. Conversely, under-reliance occurs when humans ignore correct AI advice, missing opportunities to improve performance.

To encourage appropriate reliance, prior research has primarily focused on AI-centric design interventions~\cite{effect_of_confidence}. In particular, displaying AI confidence levels has been proposed to help users gauge the likelihood of correct predictions~\cite{effect_of_confidence,vodrahalli2022uncalibrated,who_should}. Additionally, elucidating the rationale behind AI outputs through explanations remains a widely studied approach to foster appropriate reliance~\cite{effect_of_confidence,does_the,Are_explanations,li2024utilizing}.
Nonetheless, these AI-centric interventions show mixed results, indicating that they do not always lead to improved human-AI collaboration~\cite{effect_of_confidence,who_should,visual_uncertainty}.

As a result, there has been a growing focus on human self-confidence, which constitutes another key element of appropriate reliance~\cite{chong2022human,vodrahalli2022humans,are_you_really,mahmood2024designing}. Prior studies highlight the pivotal role of human DMs' self-confidence in the adoption of AI recommendations~\cite{chong2022human}.
However, self-confidence is often treated as a fixed factor, overlooking how it can be shaped to mitigate over- and under-reliance. 
\takehiro{Therefore, in this work, we propose that self-confidence shaping could serve as a viable method for enhancing human-AI team performance. }



\subsection{AI-Assisted Decision-Making in Financial Context}
\takehiro{Finance is a significant domain for AI-assisted decision-making, as evidenced by the substantial body of literature on the topic~\cite{hase-bansal-2020-evaluating,effect_of_confidence,li2024utilizing,vodrahalli2022uncalibrated}.
Its high-stakes nature, where mistakes can lead to significant losses~\cite{towards_science,biran2017human}, combined with the frequent overconfidence of DMs~\cite{barber2001boys,grevzo2021overconfidence}, makes it an ideal setting for our study on self-confidence in AI-assisted decision-making.}

\takehiro{Yet, many commonly studied financial tasks in AI-assisted decision-making, such as income prediction, fail to capture the truly high-stakes nature of finance~\cite{towards_science,li2024utilizing}. Few studies address investment decisions like stock movement prediction, which can incur substantial losses~\cite{biran2017human}, and they typically rely solely on numerical data, overlooking textual information such as corporate performance descriptions that are widely used in real-world investment scenarios~\cite{takayanagi_naacl}. Therefore, we develop a text-based stock movement prediction dataset for AI-assisted decision-making.}

\section{AI-Assisted \takehiro{Decision-Making}}
\subsection{Collaborative Decision-Making}
In this work, we study the setting of \emph{AI-assisted decision-making}. Let $\mathcal{X}$ denote the space of decision-making tasks. Assume a binary classification setting. We consider a dataset $D=\{x_i,y_i\}_{i=1}^N$, where $x_i\in\mathcal{X}$ denote the features of task $i$, and $y_i\in\{-1,+1\}$ denote the target label of task \( i \), respectively. \(N\) denotes the number of instances in the dataset.

\paragraph{Human Decision-Maker (DM)}
A human DM provides an independent judgment function \hashimoto{$h:\mathcal{X}\rightarrow\{-1,+1\}$}:
\[
  y^h \;=\; h\bigl(x;\,\theta_h\bigr),
\]
where \(y^h \in \{-1, +1\}\) is the human’s independent decision for task \( x \), and $\theta_{h}$ represents the parameters governing the human decision function.

\paragraph{AI Model}
An AI model $m:\mathcal{X}\rightarrow\{-1,+1\}$ offers a decision recommendation:
\[
  y^m \;=\; m\bigl(x;\,\theta_m\bigr),
\]
where \(y^m \in \{-1, +1\}\) is the model’s predicted label for task \( x \), and $\theta_{m}$ represents the parameters governing the AI model.

\paragraph{Human-AI Team Decision}
The human makes the final decision by incorporating their own judgment and the AI recommendation using the human-AI team decision-making function $f:\mathcal{X}\times\{-1,+1\}\times\{-1,+1\} \rightarrow \{-1,+1\}$:
\[
  d \;=\; f\bigl(x,\, m(x;\theta_m),\, h(x;\theta_h)\bigr),
\]
where \( f(\cdot) \) represents the human-AI team decision-making function to determine the final decision \(d \in \{-1, +1\}\) for task \( x \).

\paragraph{Human-AI Team Performance}
Given a dataset
\(
D = \{ (x_i, y_i) \mid 1 \le i \le N \},
\)
the goal in human-AI collaboration is to achieve a strong human-AI team performance, for example by maximizing accuracy. We can express this metric as:
 \[
\frac{1}{N} \sum_{(x_i, y_i) \in D} \mathbf{1}\Bigl\{ f\bigl(x_i,\, m(x_i;\theta_m),\, h(x_i;\theta_h)\bigr) = y_i \Bigr\},
\]
where the indicator function \(\mathbf{1}\{ \cdot \}\) returns 1 if the final decision equals \(y_i\) and 0 otherwise.

To achieve high human-AI team performance, it is essential to capture the human-AI team decision-making model \(f(\cdot)\) in a way that accurately represents the human DM’s thought process. Recent work in human-AI collaboration has increasingly emphasized human confidence when examining how humans incorporate AI recommendations into their final decisions~\cite{chong2022human,gemalmaz2022understanding,schemmer2023appropriate,wang2018does,mahmood2024designing}.
For instance, a threshold-based collaborative decision-making model is described as follows~\cite{mahmood2024designing}:
\[
f\bigl(x, m(x;\theta_m), h(x;\theta_h)\bigr) 
=
\begin{cases}
   h(x;\theta_h), & \text{if } c > \tau,\\[4pt]
   m(x;\theta_m), & \text{otherwise},
\end{cases}
\]
where $c \in \mathbb{R}$ denotes the human DM’s self-confidence, and $\tau$ is the confidence threshold: if $c \geq \tau$, the human’s decision is adopted; otherwise, the AI’s recommendation is followed.
Since self-confidence $c$ is assumed to be a fixed factor, this model does not allow us to study the process by which human DM forms self-confidence nor does it capture how shaping self-confidence can impact the human-AI team performance.
\paragraph{Self-Confidence}
To account for the self-confidence formation process, we formulate the human-AI collaborative decision-making model using a self-confidence function \(
F_\phi:\mathcal{X}\times\Theta_h\rightarrow\mathbb{R}
\):\vspace{0.5em}
{
\resizebox{\columnwidth}{!}{$
\begin{aligned}
f\bigl(x_i,\, m_i,\, h_i\bigr)=\begin{cases}
h_i, & \text{if } c_i \ge \tau,\\[1mm]
m_i, & \text{otherwise,}
\end{cases} \quad \text{with } c_i = F_\phi\bigl(x_i,\theta_h\bigr)
\end{aligned}
$}\vspace{0.5em}
}
where {\small \(f\bigl(x_i,\, m(x_i;\theta_m),\, h(x_i;\theta_h)\bigr)\)} represents the human-AI team decision for instance \(i\), \(m_i\) and \(h_i\) denote the outputs of the AI model and the human DM for instance \(i\), and \(c_i\) is the associated self-confidence. While \(f(\cdot)\) follows a threshold-based collaborative decision-making model, we integrate the self-confidence formation process by introducing a self-confidence function $F_\phi$ where \(\Theta_h\) represents the human DM's trait space.

Modeling self-confidence enables us to study self-confidence shaping. We first examine the potential of self-confidence shaping for improved human-AI team performance under the assumption of perfect control over \(c_i\) (\textbf{RQ1}). Then, we assess whether \(c_i\) can be predicted from task characteristics \(x_i\) and user traits \(\theta_h\) (\textbf{RQ2}). Finally, we consider modifying task feature \(x_i\) as an intervention to influence \(c_i\) (\textbf{RQ3}).
\section{Experiments}
\subsection{Dataset Construction}\label{sec:dataset}


\takehiro{We developed a text-based stock-movement prediction dataset using earnings conference call (ECC) transcripts. ECCs are meetings where company managers and analysts discuss operations and future plans~\cite{keith-stent-2019-modeling}, conveying critical market information that influences stock prices~\cite{ecc_influence}. Consequently, these transcripts are well-suited for our investment tasks.}

To develop our dataset, we first collected a set of short ECC summaries curated by professional journalists at Reuters, denoted as $x \in \mathcal{X}$~\cite{mukherjee-etal-2022-ectsum}. We selected 20 companies, and for each company, manually created three variant excerpts $(x_{\text{positive}}, x_{\text{neutral}}, x_{\text{negative}})$, each assigned a sentiment label $s \in \{\text{positive}, \text{neutral}, \text{negative}\}$ while preserving the factual content. We then verified that no factual errors or additional information were introduced, confirming that sentiment is the only change. 
Finally, each excerpt was anonymized by removing company names, following prior work~\cite{biran2017human}. 
This gives us a total of $20 \times 3 = 60$ excerpts, or $N = 60$.
A company's three-day stock movement is labeled as $y \in \{-1, +1\}$ (downward/upward, respectively). Companies were selected to balance upward and downward movements, with an average excerpt length of 51.2 words.

\begin{figure}
    \centering
    \includegraphics[width=.9\linewidth]{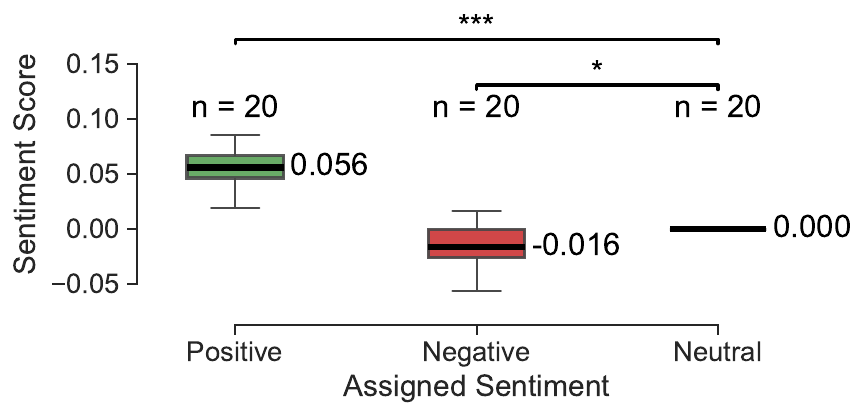}
    \caption{Box plots of sentiment scores for excerpts grouped by the assigned sentiment. Welch’s t-test indicate a highly significant difference between Positive and Neutral (***, $p\mkern-4mu<\mkern-4mu0.001$) and a marginally significant difference (*, $p\mkern-4mu<\mkern-4mu0.1$) between Negative and Neutral.}
    \label{lm_sentiment}
\end{figure}

\takehiro{To confirm the assigned sentiment, we performed financial sentiment analysis using the Loughran-McDonald dictionary~\cite{loughran2011liability}. We count the occurrences of positive and negative words from the dictionary, normalize these counts by the length of the excerpt, and then subtract the normalized count of negative words from the normalized count of positive words to obtain a sentiment score for each excerpt. As shown in Figure~\ref{lm_sentiment}, excerpts labeled positive have higher scores, while those labeled negative have lower scores, reflecting the intended sentiment.}


\subsection{Procedure}\label{sec:experimental_process}

We recruited participants via Prolific~\footnote{\url{https://www.prolific.com/}} and assigned each participant 20 financial decision-making tasks using excerpts from 20 unique companies (with randomized order and sentiment: neutral, negative, or positive). After a brief tutorial and two practice tasks to learn how to use a slider ranging from \(-100\) (decrease) to \(100\) (increase) for predictions and self-confidence, participants completed the formal tasks. 
\takehiro{In each task, participants predicted stock movement based solely on an excerpt in the first round. In the second round, they received an AI prediction (with randomly assigned stated accuracy of either 51\% or 80\%) before finalizing their decision.}

\takehiro{We measured participants' self-confidence and predictions using a slider ranging from \(-100\) to \(100\). We denote the slider values in the first and second round as \(h^{(1)}\) and \(h^{(2)}\), respectively. Human self-confidence $c$ is calculated as the absolute value of the $h^{(1)}$. Values near zero indicate low confidence, whereas values close to \(\pm 100\) indicate high confidence.}
The independent judgment \(y^h\) and final decision \(d\) were computed as follows:
\[
y^h = \begin{cases}
1, & \text{if } h^{(1)} > 0,\\[4pt]
-1, & \text{if } h^{(1)} < 0,
\end{cases}
\quad
d = \begin{cases}
1, & \text{if } h^{(2)} > 0,\\[4pt]
-1, & \text{if } h^{(2)} < 0.
\end{cases}
\]
Two attention checks were randomly inserted, and data from participants failing these were excluded.

\subsection{Participants}
In our main study, we recruited 125 UK-based native English speakers via Prolific, all of whom met a minimum 80\% approval rating and were allowed to participate only once. Participants’ ages ranged from 18 to 81 (mean = 37.7, SD = 13.2), and the median completion time was 32.5 minutes. They received a base payment of £8 per hour, with a potential bonus of £2 for achieving an accuracy above 85\%. After excluding four participants for failing the attention checks, we obtained 121 valid respondents. 
\takehiro{Prior to the main study, we ran a pilot study with five participants under identical screening criteria. The median completion time was 29.5 minutes, so we set the completion time to 30 minutes for the main study.}



\section{Evaluations}\label{sec:evaluations}

\paragraph{RQ1: Impact of Self-Confidence Shaping}
First, we aim to quantify how much human-AI team performance can improve through the deliberate shaping of human self-confidence. To capture the extent to which human DMs accept AI recommendations, we define the \takehiro{AI recommendation Acceptance Rate (AR)} as the proportion of instances where a human DM adopts the AI’s prediction when their independent judgment differs from the AI recommendation.
\[
\text{AR} = \frac{\sum_{i=1}^{N} \mathbf{1}\{y^h_i \neq y^m_i \text{ and } d_i = y^m_i\}}{\sum_{i=1}^{N} \mathbf{1}\{y^h_i \neq y^m_i\}},
\]
where, $y^h_i$ is the human's initial independent judgement,  $d_i$ is the human's final prediction, $y^m_i$ is the AI's prediction for instance $i$, \( \mathbf{1}\{\cdot\} \) denotes the indicator function, and $N$ represents the total number of instances.

\takehiro{We then perform a scenario analysis to assess the impact of shaping self-confidence on fostering appropriate reliance. Following previous work, we define inappropriate reliance as either over-reliance (when individuals align with AI predictions even though the AI is incorrect) or under-reliance (when individuals reject AI predictions even though the AI is correct)~\cite{are_you_really}.}

\vspace{1em}
\noindent \textbf{Under-Reliance Scenario:} In this scenario, highly confident individuals may ignore correct AI recommendations (underreliance).  We compute the relative change in the \takehiro{AI recommendation acceptance rate} using the following metric:
\begin{equation}
\scalebox{0.9}{$
\delta_{\text{under}}
= \frac{\text{AR}\bigl(c_i=\text{low},\, y^m_i=\text{correct}\bigr)-\text{AR}\bigl(c_i=\text{high},\, y^m_i=\text{correct}\bigr)}
       {\text{AR}\bigl(c_i=\text{high},\, y^m_i=\text{correct}\bigr)}.
$}
\end{equation}
where $c_i$ denotes the self-confidence, and  $y^m_i$ represent the AI prediction for an instance $i$. We calculate $c_i$ as the absolute value of the slider value $h^{(1)}$. 
$\text{AR}(c_i=\text{high},\, y^m_i=\text{correct})$ denotes the AR when the self-confidence is high and the AI prediction is correct.
We classify a decision as high self-confidence if its self-confidence exceeds the sample median, and low self-confidence otherwise.
The value of $\delta_{\text{under}} $ indicates a proportional change in the \takehiro{AI recommendation acceptance rate} between the high-confidence and low-confidence conditions when the AI prediction is correct.

\vspace{1em}
\noindent \textbf{Over-Reliance Scenario:} Here, low confident individuals may too readily follow incorrect AI recommendations (overreliance). We compute the relative change in the AI recommendation acceptance rate using the following metric:
\begin{equation}
\scalebox{0.9}{$
\delta_{\text{over}}
= \frac{\text{AR}\bigl(c_i=\text{high},\, y^m_i=\text{wrong}\bigr)-\text{AR}\bigl(c_i=\text{low},\, y^m_i=\text{wrong}\bigr)}
       {\text{AR}\bigl(c_i=\text{low},\, y^m_i=\text{wrong}\bigr)}.
$}
\end{equation}
The value of $\delta_{\text{over}}$ indicates a proportional change in the AI recommendation acceptance rate between the high-confidence and low-confidence conditions when the AI prediction is wrong. 

\begin{table}[t]
  \centering
  \caption{
    Acceptance rate by self-confidence (Conf.) and AI stated accuracy. 
    Superscript $^{\mathsection}$ denotes a significant difference (Welch's $t$-test, $p < 0.001$) between the two AI accuracy conditions (i.e., 51\% vs.\ 80\%), 
    and $^{\dagger}$ denotes a significant difference (Welch's $t$-test, $p < 0.001$) between the two self-confidence (i.e., Low vs.\ High).
  }
  \label{tab:advice_taking_rate}
\resizebox{0.8\linewidth}{!}{%
    \begin{tabular}{llll}
      \toprule
      \multirow{2}{*}{\textbf{Conf.}} 
        & \multicolumn{2}{c}{\textbf{AI Stated Acc.}} & \multirow{2}{*}{\textbf{Avg.}}\\
      \cmidrule(lr){2-3}
       & Low (51\%) & High (80\%) \\
      \midrule
    High   & $0.160^{\mathsection\dagger}$ & $0.298^{\dagger}$ & $0.232^{\dagger}$ \\
    Low    & $0.398$ & $0.449$ & $0.423$ \\
    \midrule
    Avg.& $0.269^{\mathsection}$ & $0.363$ & $0.317$ \\
      \bottomrule
    \end{tabular}
  }
\end{table}

\paragraph{RQ2: Self-Confidence Prediction}
\takehiro{We then investigate the extent to which self-confidence can be predicted using user traits and task characteristics. An accurate predictive model allows us to determine when self-confidence deviates from the optimal level, thereby indicating when the intervention is needed.}

In our study, we treat the prediction task as a binary classification problem where the outcome variable is whether a DM’s self-confidence is high or low. We define high confidence as any confidence score above the median value in our training data and low confidence as any score below the median. The predictors include user traits (such as demographic information and financial traits) and task characteristics (such as assigned sentiment and a pre-trained BERT-based text representation of the task description~\cite{devlin-etal-2019-bert}).

Accuracy is our primary evaluation metric. 
\takehiro{We randomly split the dataset into training, validation, and test sets in a 50:25:25 ratio, respectively.
We perform 5-fold cross-validation to select the best model, which is then evaluated on the test set.}
We repeat this over 10 seeds and report the average accuracy. Our evaluated models include a rule-based approach, which predicts high confidence for tasks with positive or negative sentiment and low confidence for neutral tasks, as well as machine learning models such as logistic regression, a multilayer perceptron (MLP), and LightGBM.

\paragraph{RQ3: The Relationship Between Sentiment and Self-Confidence}
Finally, we evaluate the association between task features \(x_i\) and a decision maker's self-confidence. We posit that self-confidence is influenced both by objective information (e.g., facts) and by qualitative impressions (e.g., sentiments) present in the task features~\cite{griffin1992weighing}. In our experiment, we focus on how the sentiment in a text is associated with self-confidence. We exclusively focus solely on sentiment because altering factual content may compromise objectivity and trustworthiness.

In particular, we compare self-confidence across excerpts with different assigned sentiments and test for significant differences between positive/negative and neutral excerpts.
Then, to examine how the intensity of a text’s sentiment relates to self-confidence, we compute Pearson’s correlation between the self-confidence and the absolute value of sentiment score (see Section~\ref{sec:dataset}). Since we want to focus on sentiment intensity, we take the absolute value of the sentiment score.


\section{Results}

\begin{table}[t]
  \centering
  \caption{Acceptance rate (AR) and relative rate change by scenario. This table shows the AR under high and low confidence conditions and the corresponding relative rate changes for under- and over-reliance scenarios.}
  \label{tab:AR}
  \begin{tabular}{lccl}
    \toprule
    \multirow{2}{*}{\textbf{Scenario}} & \multicolumn{2}{c}{\textbf{Confidence}} & \multirow{2}{*}{\textbf{Rate Change} ($\delta$)} \\
    \cmidrule(r){2-3}
                      & High & Low & \\ 
    \midrule
    Under-Reliance & 0.279 & 0.412 & $ +47.6\%$ ($\delta_
    {\text{over}}$) \\
    Over-Reliance & 0.209 & 0.432 & $-51.6\%$  ($\delta_
    {\text{under}}$)\\
    \bottomrule
  \end{tabular}
\end{table}

\subsection{RQ1: Impact of Self-Confidence Shaping}\label{result:rq1}
We begin by examining how human-AI team performance can be improved by deliberately \takehiro{shaping} the \takehiro{self-confidence} of human DMs.
\takehiro{The acceptance rate, our primary metric, measures how often human DMs follow AI recommendations. We compare it across self-confidence (high vs.\ low) and stated AI prediction accuracy (high vs.\ low) to assess their influence on acceptance behavior.}

Table~\ref{tab:advice_taking_rate} shows the acceptance rate by self-confidence and AI-stated accuracy. Confidence is classified as high or low using the median, and AI accuracy is randomly set to 80\% or 51\% per task (see Section~\ref{sec:experimental_process}).
First, we can observe that human DMs tend to follow AI recommendations less frequently when they are highly confident in their own judgments. High-confidence instances show significantly lower acceptance rates than low-confidence ones (\takehiro{$0.160$ vs.\  $0.398$, $0.298$ vs.\ $0.449$, and $0.232$ vs.\ $0.423$, indicated by \(^{\dagger}\)}). Moreover, acceptance increases with higher AI-stated accuracy \takehiro{($0.269$ vs.\ $0.363$, indicated by \(^{\mathsection}\))} on average. High-confidence users are significantly more inclined to follow AI recommendations at 80\% accuracy \takehiro{($0.160$ vs.\ $0.298$, indicated by \(^{\mathsection}\))}, while low-confidence users show no significant difference between accuracy levels \takehiro{($0.398$ vs.\ $0.449$)}.

Interestingly, even when the AI’s accuracy is as high as 80\%, highly confident individuals remain relatively reluctant to follow AI recommendations, exhibiting an acceptance rate of 29.8\%. In contrast, low-confidence DMs, despite encountering a lower AI accuracy of 51\%, demonstrate a higher acceptance rate of 39.8\%. This observation underscores the critical role of self-confidence: regardless of the AI’s stated accuracy, individuals’ self-confidence significantly influence how likely they are to follow AI recommendations.

Next, we perform a scenario analysis to assess the impact of shaping self-confidence under two scenarios: under-reliance and over-reliance scenario. In under-reliance scenario, DMs ignore correct AI recommendations. Lowering self-confidence in this case is expected to encourage a more appropriate reliance, measured by \(\delta_{\text{under}}\). Here, a positive value indicates enhanced adoption of accurate AI recommendations and a corresponding gain in human-AI team performance. Conversely, in the over-reliance scenario, DMs are prone to accepting erroneous AI recommendations. Increasing self-confidence should help reduce this erroneous reliance, quantified by \(\delta_{\text{over}}\). 
A negative value reflects a reduction in wrong advice adoption and thus indicates a performance gain.

\begin{table}[t]
\centering
\caption{Comparison of model accuracy (in \%) across 10 seeds. 
We use Welch's t-test with $p\mkern-4mu<\mkern-4mu0.001$ for statistical significance: 
$^{\dagger}$ indicates significant improvement over the rule-based approach, 
$^{\mathsection}$ indicates significant improvement over using only user traits (U), 
and $^{\mathparagraph}$ indicates significant improvement over using only task traits (T). ($\downarrow$) indicates how much the accuracy decreased for either (T) or (U), compared to (T+U).}
\label{tab:accuracy}
\resizebox{1.0\linewidth}{!}{%
\begin{tabular}{lcclll}
\toprule
      & Random           & Rule             & LightGBM                                 & LR                                      & MLP                                      \\
\midrule
$w$/T+U   & $50.0$           & $55.4$           & $66.4^{\dagger\mathsection\mathparagraph}$ & $66.9^{\dagger\mathsection\mathparagraph}$ & $67.3^{\dagger\mathsection\mathparagraph}$ \\
$w$/U     & $50.0$           & $55.4$           & $63.7^{\downarrow{4.1\%}}$                                    & $62.0^{\downarrow{7.3\%}}$                                    & $63.2^{\downarrow{6.1\%}}$                                    \\
$w$/T     & $50.0$           & $55.4$           & $63.3^{\downarrow{4.7\%}}$                                    & $62.6^{\downarrow{6.4\%}}$                                    & $63.1^{\downarrow{6.2\%}}$                                    \\
\bottomrule
\end{tabular}
}
\end{table}
Table~\ref{tab:AR} presents the results of the scenario analysis, showing the acceptance rates and corresponding relative rate changes.
We compute the relative change for under-reliance scenario as $\delta_{\text{under}} = (0.412 - 0.279) / 0.279 \approx +47.6\%$. This positive value indicates that lowering self-confidence from high to low could potentially increase the human-AI team performance by $47.6\%$, as it promotes a more appropriate reliance on the AI recommendations.
Similarly, the relative change for over-reliance is given by $\delta_{\text{over}} = (0.209 - 0.432) / 0.432 \approx -51.6\%$. Although the computed percentage is negative, this outcome is desirable because it reflects a $51.6\%$ reduction in the tendency to adopt incorrect AI recommendations. In other words, raising self-confidence from low to high helps discourage following incorrect AI predictions, thus improving team performance by $51.6\%$.
These analyses suggest that by appropriately shaping self-confidence, we can significantly mitigate both under-reliance and over-reliance on AI recommendations, thereby improving human-AI team performance.

\textbf{To answer RQ1,} our results confirm that self-confidence shaping can significantly improve human-AI team performance by reducing both over- and under-reliance. We show that human self-confidence is key to AI recommendation acceptance, and our scenario analysis indicates that perfect control over self-confidence could yield performance gains of 47.6\% and 51.6\% by mitigating under- and over-reliance, respectively. While our scenario analysis shows the potential of self-confidence shaping to enhance human-AI team performance, its assumption of perfect control over human confidence is unrealistic. In the following analysis, we evaluate the feasibility of such interventions in practice.

\subsection{RQ2: Self-Confidence Prediction}\label{result:rq2}
Having shown that self-confidence shaping can enhance human-AI collaboration, we now examine the feasibility of such interventions.
In particular, we examine the extent to which self-confidence can be predicted.

\begin{figure}
    \centering
    \includegraphics[width=\linewidth]{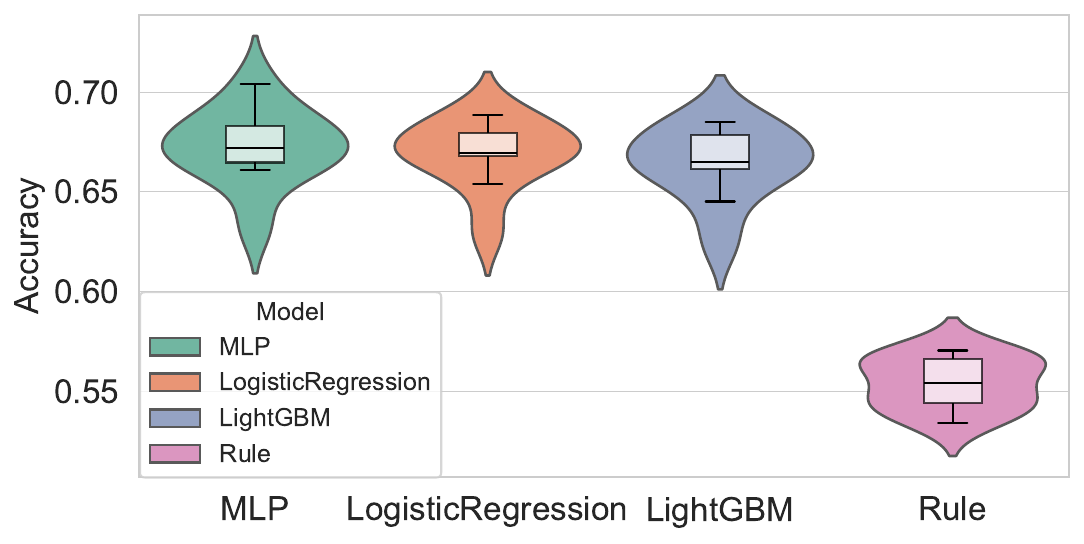}
    \caption{Violin plots showing the distribution of accuracy for machine learning and rule-based approaches across 10 seeds. Each violin depicts the kernel density of accuracy values. The box within indicates the interquartile range and the horizontal line marks the median. }
    \label{fig:box_violin_accuracy}
\end{figure}

Table~\ref{tab:accuracy} reports the average accuracy of our machine learning models and the rule-based approach, evaluated with and without the incorporation of user and task traits. Here, MLP and LR denote multilayer perceptrons and logistic regression models, respectively. Figure~\ref{fig:box_violin_accuracy} displays a violin plot of the accuracy distributions obtained from 10 random seeds when user and task traits are included.


From Table~\ref{tab:accuracy}, we observe that the rule-based approach achieves an accuracy of 55.4\%, which exceeds the random-guess baseline of 50\%. This suggests that the presence of sentiment is linked to self-confidence. Moreover, Table~\ref{tab:accuracy} shows that simple machine learning models significantly outperform the rule-based approach, with LightGBM, logistic regression, and MLP achieving accuracy improvements of 19.8\%, 20.8\%, and 21.5\%, respectively. 
\takehiro{This implies that self-confidence exhibits predictable patterns rather than being entirely random or unpredictable.}
The violin plots in Figure~\ref{fig:box_violin_accuracy} further confirm that, across 10 seeds, the machine learning models consistently outperform the rule-based approach. 

Furthermore, Table~\ref{tab:accuracy} shows that removing either user traits or task traits leads to a significant reduction in accuracy. For example, with the MLP model, accuracy decreases by 6.1\% when task traits are removed and by 6.2\% when user traits are removed. This indicates that both user and task traits play an important role in predicting self-confidence.


\begin{figure}
    \centering
    \includegraphics[width=0.9\linewidth]{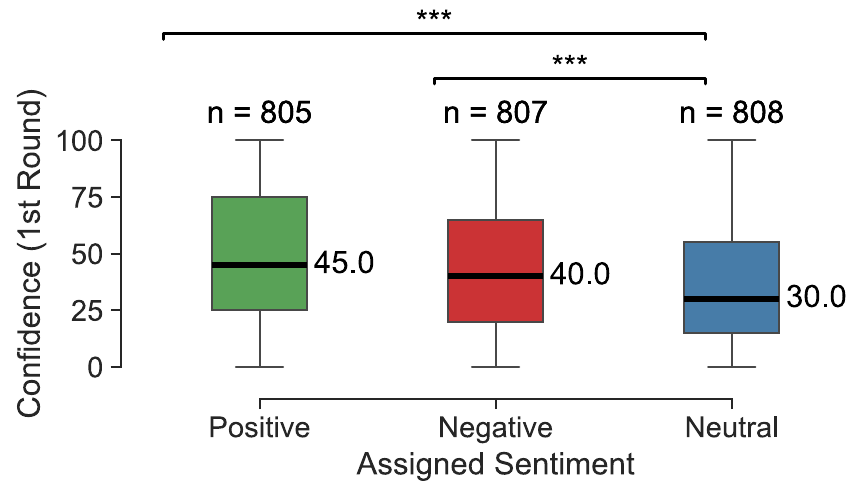}
    \caption{Box plots of self-confidence $c$ for the three task‐sentiment variants. Welch’s t-test indicates a highly significant difference between Positive and Neutral (***, $p\mkern-4mu<\mkern-4mu0.001$) and a highly significant difference (***, $p\mkern-4mu<\mkern-4mu0.001$) between Negative and Neutral.}
    \label{fig:confidence_sentiment_plot}
\end{figure}

\textbf{To answer RQ2}, our results show that we can predict high vs.\ low confidence in decision-making with about 67\% accuracy using simple machine learning models that incorporate user and task traits. These models significantly outperform the rule-based approach (55.4\%) and a random-guess baseline (50\%). Our findings also confirm the importance of both user and task traits for accurate predictions. 
These results indicate that self-confidence prediction is indeed feasible and not merely a random exercise.
Future work may explore refining or expanding user and task traits or developing more advanced modeling techniques to achieve higher accuracy.

\subsection{RQ3: The Relationship Between Sentiment and Self-Confidence}\label{result:rq3}
We now examine our hypothesis that sentiment in a text is associated with human decision-making self-confidence, serving as a preliminary step toward modifying text to shape self-confidence.


Figure~\ref{fig:confidence_sentiment_plot} shows self-confidence grouped by assigned sentiment. We observe significant differences when comparing positive vs.\ neutral and negative vs.\ neutral sentiments, indicating a relationship between sentiment and self-confidence. We also find that self-confidence is higher for tasks with positive sentiments than for those with negative sentiments. We believe this is because the intensity of positive sentiment is higher. As shown in Figure~\ref{lm_sentiment}, the absolute sentiment score is greater for positive than for negative sentiments, suggesting that the intensity of the sentiment influences self-confidence.

Then, to investigate the relationship between the intensity of the sentiment and the level of self-confidence. We conduct a correlation analysis between the absolute value of sentiment score and self-confidence. We found a weak positive Pearson correlation of 0.221 ($p\mkern-4mu<\mkern-4mu0.001$), indicating that higher sentiment intensity is generally associated with higher self-confidence. 
However, the modest strength of this linear relationship suggests that factors beyond sentiment may affect a DM’s self-confidence. For instance, expressions of uncertainty (e.g., \emphasize{it will definitely rise} vs. \emphasize{it might rise}) could substantially alter how confident DMs feel. Therefore, to develop effective text modifications for shaping self-confidence, further research is needed to investigate the association between diverse linguistic patterns, such as uncertainty, and self-confidence


\textbf{To answer RQ3,} our results demonstrate that the sentiments are significantly related to the self-confidence in human decision-making. Our findings also suggest that the intensity of the sentiment is associated with self-confidence. However, a weak positive Pearson correlation ($r=0.221,  p\mkern-4mu<\mkern-4mu0.001$) between the sentiment score and the self-confidence indicates that factors beyond sentiment affect self-confidence. Future work should investigate a broader range of linguistic patterns in a text to enable more refined self-confidence shaping.

\subsection{\takehiro{Real-World Applications}}
\takehiro{Our findings demonstrate that self-confidence shaping can mitigate both over- and under-reliance on AI recommendations, and that such interventions are feasible through self-confidence prediction and text sentiment modification. These results have implications for high-stakes decision-making. For instance, consider a financial setting where updated ECC summaries convey information about recent extreme market events, and investors need to predict whether a stock price will increase or decrease. In such cases, AI models may yield low prediction accuracy in the face of extreme events (e.g., 55\%), while experienced financial experts might achieve higher accuracy (e.g., 65\%). However, the volatility of these events can undermine experts’ self-confidence, causing them to over-rely on the less accurate AI predictions. By employing our self-confidence prediction model to detect low self-confidence in advance and modifying the ECC summaries’ sentiment to increase experts' self-confidence, the system can help experts rely on their own judgment, thus mitigating the under-reliance.}

\section{Conclusion}
This paper introduces a \takehiro{human self-confidence shaping} paradigm to enhance human-AI team performance.
First, this paper quantitatively assessed the extent to which human self-confidence shaping can improve human-AI team performance. Our scenario analysis indicated that, assuming successful interventions, self-confidence shaping mitigates both under-reliance and over-reliance on AI recommendations, potentially yielding performance gains of up to 47.6\% and 51.6\%, respectively.
Furthermore, we demonstrate the promising plausibility of such interventions: simple machine learning models achieve 67\% accuracy in predicting self-confidence, and our results indicate that modifying sentiment could be a viable method for shaping self-confidence.

While our analysis provides an initial step towards improving human-AI team performance via self-confidence shaping, significant gaps remain in implementing such interventions in real-world settings. In particular, future research should focus on improving the accuracy of self-confidence prediction by employing more sophisticated models and expanding the range of user and task traits considered. Also, further exploration of the relationship between self-confidence and linguistic patterns is needed for more refined interventions. 
By integrating human factors into AI-assisted decision-making, we offer insights into designing \takehiro{human-AI collaboration} that better support and complement human DMs.

\bibliographystyle{named}
\bibliography{ijcai25}

\end{document}